\documentclass[aps,prd,twocolumn,10pt]{revtex4}
\usepackage[colorlinks,linkcolor={red},citecolor={blue}]{hyperref}
\usepackage{eurosym}
\usepackage{graphicx}
\usepackage{amsmath}
\usepackage{amssymb}
\usepackage{amsfonts}
\usepackage{bm}
\usepackage{float}
\usepackage{color}

\def\be{\begin{equation}}
\def\ee{\end{equation}}
\def\bea{\begin{eqnarray}}
\def\eea{\end{eqnarray}}

\newcommand{\mc}[1]{\mathcal{#1}}

\newcommand{\f}[2]{\frac{#1}{#2}}

\begin{document}

\title{Non-minimal Energy-momentum squared gravity}
\author{Shahab Shahidi}
\email{s.shahidi@du.ac.ir}
\affiliation{School of Physics, Damghan University, Damghan 41167-36716, Iran.}
\date{\today}
\begin{abstract}
We consider a gravitational theory with an additional non-minimal coupling between baryonic matter fields and geometry. The coupling is second order in the energy momentum tensor and can be seen as a generalization of the energy-momentum squared gravity model. We will add a constraint through a Lagrange multiplier to ensure the conservation of the energy-momentum tensor. Background cosmological implications together with its dynamical system analysis will be investigated in details. Also we will consider the growth of matter perturbation at first order, and estimate the model parameter from observations on $H$ and also $f\sigma_8$. We will show that the model parameter should be small and positive in 2$\sigma$ confidence interval. The theory is shown to be in a good agreement with observational data.  
\end{abstract}

%\pacs{}
\maketitle
\section{Introduction}
Among many interesting foundations of the modern cosmology, one of the most controversial of them is the late time acceleration of the Universe \cite{acceleration}. The traditional solution to this observation is reconsidering the cosmological constant term in the Einstein-Hilbert theory \cite{cc}. However, the cosmological constant suffers from phenomenological/theoretical issues, persuading modern cosmologists to consider modified gravitational theories \cite{review}. There are three main streams on modifying gravity at large scales, one is considering some extra degrees of freedom, such as scalar field \cite{scalar} or a vector field \cite{proca}, and the other is to modify the gravitational interaction itself, such as massive gravity \cite{massivegravity}, Weyl-Cartan theories \cite{weylcartan}, etc. The third category is to modify the gravity action to contain higher order term in curvature tensor, such as $f(R)$ theories \cite{fR}.

There is also another line of thought about modifying the Einstein-Hilbert theory, which contains non-minimal interactions between matter field and curvature. As a very simple example, one can add a non-minimal coupling between the Ricci scalar and the matter Lagrangian \cite{fRLm}
\begin{align}
S=\int d^4x\sqrt{-g}\big[\kappa^2(R-2\Lambda)+f(R,\mc{L}_m)+\mc{L}_m\big].
\end{align}
There are however other possible choices which contains $f(R,T)$ or $f(R,T,R_{\mu\nu}T^{\mu\nu})$ gravities where $T$ is the trace of the energy-momentum tensor \cite{fRT,fRTRT}.
Cosmological consequences as well as other concepts of these theories is vastly investigated in the literature. It should be noted that in all of the above theories, the consequence of adding an extra term to the Einstein-Hilbert action is to weaken the role of the cosmological constant, or at an idealized level to nullify its effect.

Recently, a subclass of the non-minimal coupling between matter and geometry attracts some interests, which contains higher order terms in the energy momentum tensor. The action of this theory can be written as \cite{emsg}
\begin{align}
S=\int d^4x\sqrt{-g}\big[\kappa^2(R-2\Lambda)+\alpha(T_{\mu\nu}T^{\mu\nu})^\eta+\mc{L}_m\big],
\end{align}
where $\eta$ and $\alpha$ are constants. It is worth mentioning that the special case $\eta=2$ is dubbed energy-momentum squared gravity (EMSG). For small and negative values of $\eta$, the model can satisfy the observational data, at least at background level, without any cosmological constant term. However, it is shown that for other values of the parameter $\eta$ some non-zero cosmological constant, or another dynamical field is needed \cite{emsg}.

Another interesting possibility on the non-minimal coupling between matter and geometry would be to consider derivative terms containing matter fields \cite{harko,ours}. If one noted that the matter Lagrangian $\mc{L}_m$ is a scalar field and add some Galileon-like terms to the Einstein-Hilbert action, one would get the accelerated expansion of the Universe without the need for cosmological constant \cite{ours}. Also, at the background level, such models satisfy the conservation equation for the energy-momentum tensor, which is promising since other non-minimal matter couplings could not fulfill such a property. The simplest action in this line could be written as
\begin{align}
S=\int d\,^4x\sqrt{-g}\bigg[\kappa^2 R-\alpha\,\nabla_\mu f\,\nabla^\mu f\bigg]+S_m,
\end{align}
where $f=f(\mc{L}_m)$ is an arbitrary function of the matter Lagrangian. It is shown that one can obtain an accelerated expansion of the Universe for power law function $f\propto (\mc{L}_m)^\eta$, with small and negative $\eta$ and positive values of $\alpha$.

In this paper, we will explore a new possibility on the non-minimal coupling of matter and geometry. This contains a coupling between the Ricci tensor and the energy-momentum tensor in the form
\begin{align}\label{term}
\mc{L}\supseteq R_{\mu\nu}T^{\alpha\mu}T_\alpha^{~\nu}.
\end{align}
The above interaction term can be seen as a generalization of the EMSG theory \cite{emsg}. This is true because one can write the interaction term of the EMSG as
$$\alpha g_{\mu\nu}T^{\alpha\mu}T_\alpha^{~\nu}.$$
Now considering the fact that in the maximally symmetric space-times we have $R_{\mu\nu}=\alpha g_{\mu\nu}$, one can write the interaction term of the EMSG in this space-times as 
$$R_{\mu\nu}T^{\alpha\mu}T_\alpha^{~\nu}.$$
The interaction term we are considering in this paper can then be considered as a generalization of the EMSG interaction terms to general space-times. There is also a difference between this term and the term $RT^2$ where $T$ is the trace of the energy-momentum tensor. The later term is a subset of $f(R,T)$ gravity, which our interaction term does not belong to it.

As we have discussed earlier, the interaction term \eqref{term} does not satisfy the conservation of the energy-momentum tensor. In this paper, we will impose the conservation of the energy-momentum tensor by adding a Lagrange multiplier term to the action. The resulting theory would have the effects of the non-minimal matter/geometry couplings in a way that the energy-momentum itself is conserved.

The paper is organized as follows. In the next section we introduce the model and obtain the field equations. In section \ref{cos} we will consider the background cosmology of the model and in section \ref{dyn} we investigate the dynamical system analysis of the model. In section \ref{pert} we obtain the dynamical equation for the matter density contrast at first order in perturbation variables. We then use the observational data on the Hubble parameter $H$ and also for $f\sigma_8$ to estimate the best fit values of the model parameter. We conclude in section \ref{conc}.
\section{The model}
Let us consider the action functional of the form
\begin{align}\label{action}
S=\int d^4x&\sqrt{-g}\big[\kappa^2(R-2\Lambda)\nonumber\\&+\alpha R_{\mu\nu}T^{\alpha\mu}T_\alpha^{~\nu}+A_\mu\nabla_\nu T^{\mu\nu}+\mc{L}_m\big],
\end{align}
where $\Lambda$ is the cosmological constant, $\alpha$ is an arbitrary constant with mass dimension $M^{-6}$, and $A_\mu$ is a Lagrange multiplier ensuring that the theory respects energy-momentum conservation. 
It should be noted that, there are strong constraints on the conservation of the matter sources. Here, we have alleviate this problem by introducing a Lagrange multiplier term to insure that the energy-momentum tensor remains conserved. The same procedure have been done before, for example in $f(R,T)$ theories \cite{fRTRT}. One can then consider the vector $A_\mu$ as a modulator potential which determines the required force we need to keep the matter conserved and not convert directly to geometry.

Varying the action \eqref{action} with respect to the metric gives
\begin{widetext}
\begin{align}\label{eq1}
\kappa^2(G_{\mu\nu}&+\Lambda g_{\mu\nu})=\f12T_{\mu\nu}+\f12g_{\mu\nu}(\mc{L}_m g_{\alpha\beta}-T_{\alpha\beta})\nabla^\alpha A^\beta+\f12A^\alpha\nabla_\alpha T_{\mu\nu}+T_{\alpha(\mu}\nabla_{\nu)}A^\alpha\nonumber\\&+\alpha g_{\mu\nu}\bigg(T^{\alpha\beta}\nabla_\beta\nabla_\gamma T_\alpha^{~\gamma}+\f12\nabla_\beta T_{\alpha\gamma}\nabla^\gamma T^{\alpha\beta}+R_{\alpha\gamma\beta\sigma}\Big(\mc{L}_m g^{\gamma\sigma}T^{\alpha\beta}-\f12T^{\alpha\beta}T^{\gamma\sigma}\Big)\bigg)\nonumber\\
&+\alpha\Big(R^\alpha_{~(\mu}T_{\nu)}^{~\beta}T_{\alpha\beta}-2\mc{L}_m R^\alpha_{~(\mu}T_{\nu)\alpha}-R_{\alpha\beta\gamma(\mu}T_{\nu)}^{~\alpha}T^{\beta\gamma}+R^{\alpha\beta}(T_{\alpha\mu}T_{\beta\nu}-T_{\alpha\beta}T_{\mu\nu})\nonumber\\&+T^\alpha_{~(\mu}\Box T_{\nu)\alpha}-T^{\alpha\beta}\nabla_\beta\nabla_{(\mu}T_{\nu)\alpha}+\nabla_\alpha T_{\mu\beta}\nabla_\alpha T_\nu^{~\beta}-\nabla^\beta T^\alpha_{~(\mu}\nabla_{\nu)}T_{\alpha\beta}\Big).
\end{align}
Also, taking the covariant divergence of the equation of motion \eqref{eq1} will determine the equation of motion of the Lagrange multiplier $A_\mu$
\begin{align}\label{cov1}
\f12&\mc{L}_m A^\alpha R_{\alpha\nu}-\f12A^\alpha R_\alpha^{~\beta}T_{\nu\beta}+\f12\mc{L}_m\Box A_\nu-\f12T_{\nu\alpha}\Box A^\alpha+\f12\nabla\alpha A^\beta(\nabla_\nu T_{\alpha\beta}-\nabla_\alpha T_{\nu\beta}-\nabla_\beta T_{\nu\alpha})+\nabla_{[\alpha}\mc{L}_m\nabla_{\nu]}A^\alpha\nonumber\\&+\alpha\Big[T^{\beta\gamma}T_\nu^{~\alpha}\nabla_\alpha R_{\beta\gamma}+2T_{\beta(\alpha}R_{\nu)}^\beta\nabla^\alpha \mc{L}_m+\f12T_\nu^{~\beta}\nabla^\alpha R(\mc{L}_m g_{\alpha\beta}-T_{\alpha\beta})+\mc{L}_mT^{\alpha\beta}\nabla_\beta R_{\nu\alpha}+\mc{L}_mR^{\alpha\beta}\nabla_\beta T_{\nu\alpha}\nonumber\\&-T^{\beta\gamma}T_\nu^{~\alpha}\nabla_\gamma R_{\alpha\beta}+2R^{\alpha\beta}\nabla_\gamma T_{\beta[\alpha}T_{\nu]}^{~\gamma}-\f12T^{\alpha\beta}T^{\gamma\delta}(\nabla_\nu R_{\alpha\gamma\beta\delta}+2\nabla_\delta R_{\nu\alpha\beta\gamma})-\nabla_\nu(R_{\alpha\beta}T^{\alpha\beta}\mc{L}_m)\Big]=0.
\end{align}
\end{widetext}
Varying the action \eqref{action} with respect to the Lagrange multiplier $A_\mu$ gives
\begin{align}\label{cons}
	\nabla_\mu T^{\mu\nu}=0.
\end{align}

One can write the metric field equation as
\begin{align}
\kappa^2(G_{\mu\nu}+\Lambda_{eff}g_{\mu\nu})=\f12(T_{\mu\nu}+T^{eff}_{\mu\nu}),
\end{align}
where we have defined the effective cosmological constant and energy-momentum tensor as
\begin{align}\label{coscon}
\kappa^2\Lambda_{eff}&=\kappa^2\Lambda+\f12(\mc{L}_m g_{\alpha\beta}-T_{\alpha\beta})\nabla^\alpha A^\beta\nonumber\\&+\alpha\bigg(T^{\alpha\beta}\nabla_\beta\nabla_\gamma T_\alpha^{~\gamma}+\f12\nabla_\beta T_{\alpha\gamma}\nabla^\gamma T^{\alpha\beta}\nonumber\\&+R_{\alpha\gamma\beta\sigma}\Big(\mc{L}_m g^{\gamma\sigma}T^{\alpha\beta}-\f12T^{\alpha\beta}T^{\gamma\sigma}\Big)\bigg)
\end{align}
and
\begin{align}
T_{\mu\nu}&^{eff}=A^\alpha\nabla_\alpha T_{\mu\nu}+2T_{\alpha(\mu}\nabla_{\nu)}A^\alpha\nonumber\\&+2\alpha\Big(R^\alpha_{~(\mu}T_{\nu)}^{~\beta}T_{\alpha\beta}-2\mc{L}_m R^\alpha_{~(\mu}T_{\nu)\alpha}-R_{\alpha\beta\gamma(\mu}T_{\nu)}^{~\alpha}T^{\beta\gamma}\nonumber\\&+R^{\alpha\beta}(T_{\alpha\mu}T_{\beta\nu}-T_{\alpha\beta}T_{\mu\nu})-T^{\alpha\beta}\nabla_\beta\nabla_{(\mu}T_{\nu)\alpha}\nonumber\\&+T^\alpha_{~(\mu}\Box T_{\nu)\alpha}+\nabla_\alpha T_{\mu\beta}\nabla_\alpha T_\nu^{~\beta}-\nabla^\beta T^\alpha_{~(\mu}\nabla_{\nu)}T_{\alpha\beta}\Big).
\end{align}
As can be seen from the above relations, the effective energy-momentum tensor and the effective cosmological constant, have contributions from the non-minimal interactions between the Lagrange multiplier $A_{\mu}$ and the matter field, and also from the non-minimal interactions of the matter field and geometry. Also, it should be noted that the effective energy-momentum tensor includes derivative couplings of the matter sources. This is a generic property of theories with non-minimal matter-geometry couplings \cite{fRT} and also theories with derivative matter couplings \cite{ours}.
\section{Cosmological consequences}\label{cos}
Let us assume that the Universe is described by the flat FRW metric
\begin{equation}
ds^{2}=a^{2}(-dt^{2}+d\vec{x}^2),
\end{equation}%
where $a=a(t)$ is the scale factor and $t$ is the conformal time. We define the Hubble parameter as $H=\dot{a}/a$, describing the rate of expansion of the Universe.
Here dot denotes derivative with respect to $t$.

Let us also assume that the Universe is filled with a perfect fluid, with
Lagrangian density $\mc{L}_{m}=-\rho $ and the energy-momentum tensor
\begin{equation}
T_{\nu }^{\mu }=\mathrm{diag}\left( -\rho ,p,p,p\right),
\end{equation}
where $\rho$ is the energy density of the baryonic matter and $p$ is its thermodynamics pressure.
With the above assumptions, the energy-momentum conservation equation \eqref{cons} takes the form
\begin{align}\label{cons2}
\dot{\rho}+3H(\rho+p)=0.
\end{align}
For the Lagrange multiplier, we choose $A_{\mu}=(A_0(t),\vec{0})$. It should be noted that the Lagrange multiplier is a vector field over FRW space-time. As a result the isotropy and homogeneity of the space-time implies that the vector field $A_{\mu}$ has only a non-zero temporal component in Cartesian coordinates.

One can then obtain the Friedmann and Raychaudhuri equations as
\begin{align}\label{frid1}
6\kappa^2 H^2=a^2(\rho+2\kappa^2\Lambda)+6\alpha H(\rho+p)\dot\rho,
\end{align}
and
\begin{align}\label{ray1}
4\kappa^2(\dot{H}-H^2)=&-a^2(\rho+p)+A_0\big(\dot{p}-H(\rho+p)\big)\nonumber\\&-8Hp\dot{p}+2(\dot{H}+2H^2)(p^2-\rho^2)\nonumber\\&-4(\dot{H}-H^2)p\rho+\f{d^2}{dt^2}(p^2+\rho^2).
\end{align}
The covariant divergence of the metric field equation \eqref{cov1} gives
\begin{align}
\left(\f32HA_0-3\alpha p(2H^2+\dot{H})\right)\Big(\dot\rho+\dot{p}+2H(\rho+p)\Big)=0.
\end{align}
\begin{figure*}
	\includegraphics[scale=0.7]{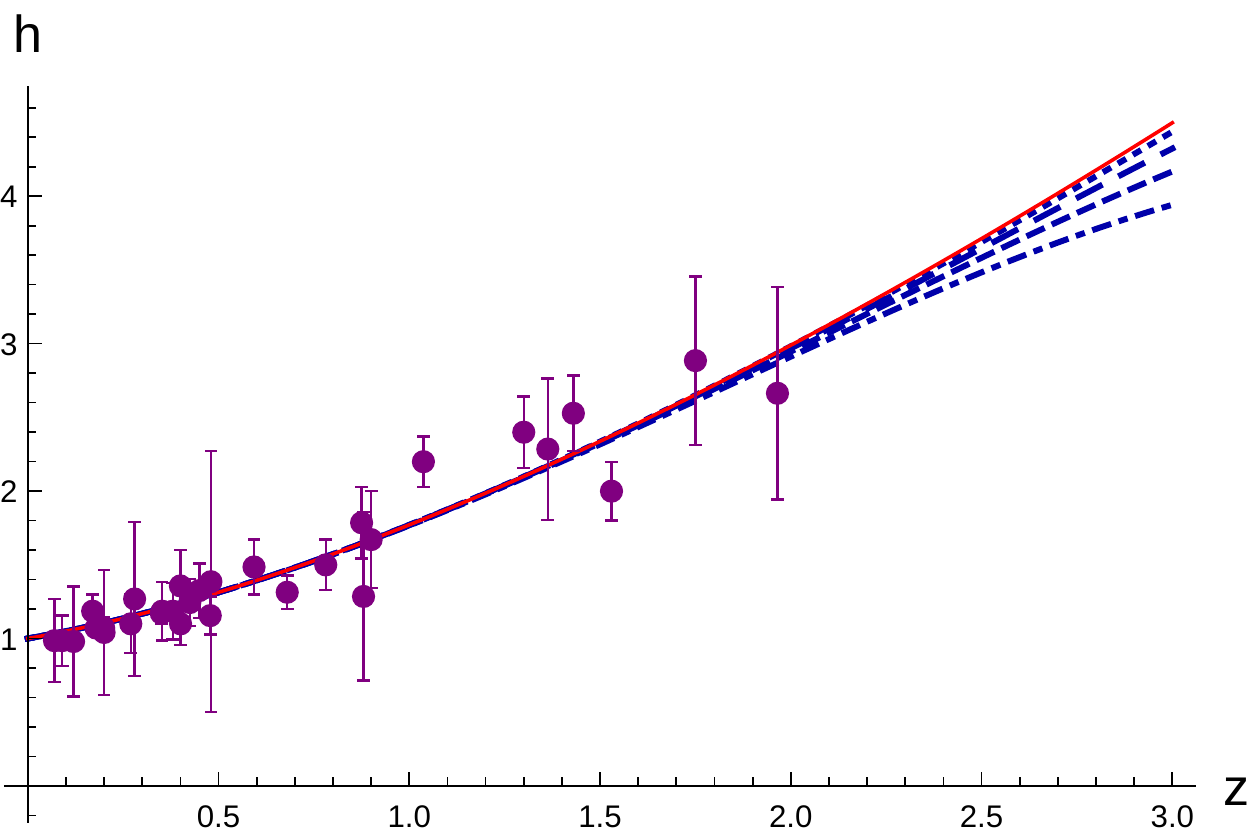}
	\includegraphics[scale=0.7]{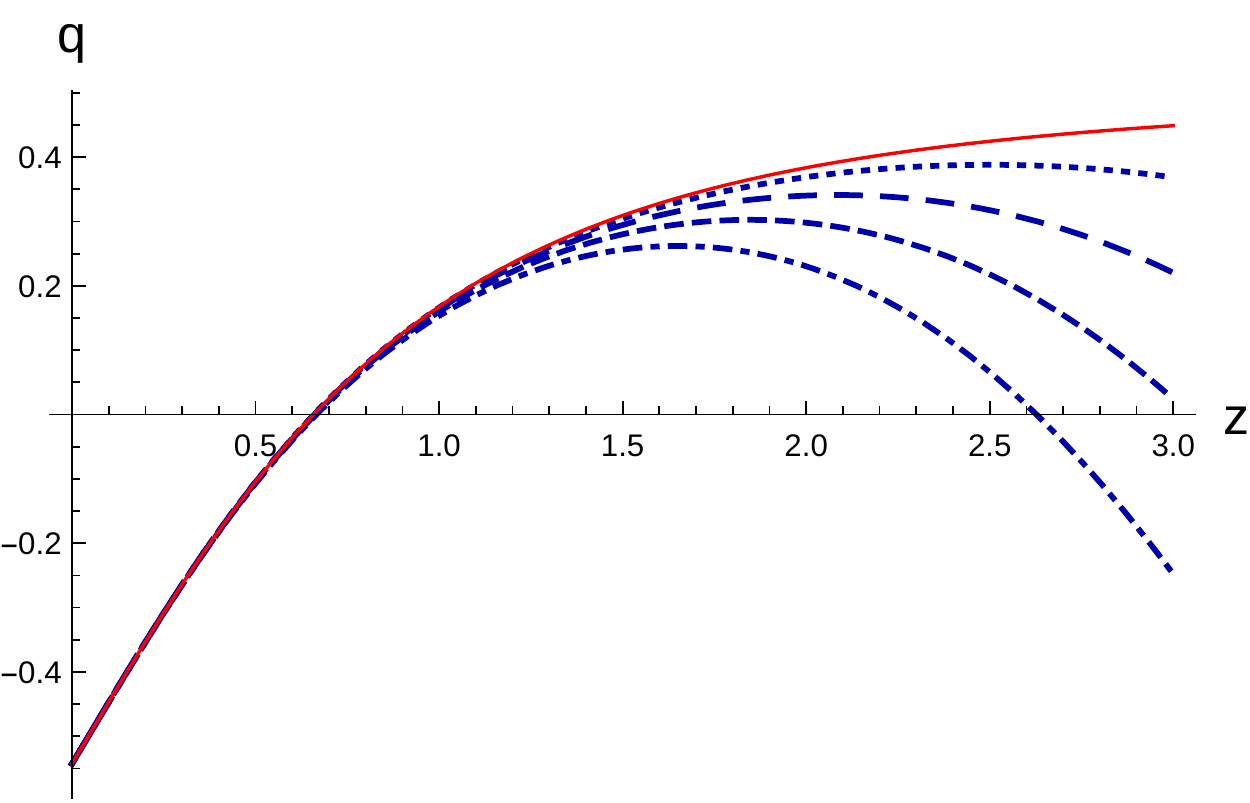}
	\caption{The Hubble and deceleration parameters as a function of the redshift $z$ for $\beta\times10^{5}=(0.1,0.3,0.6,1.1)$ indicated as dotted, long-dashed, dashed and dot-dashed respectively. The $\Lambda$CDM curve is depicted as a red curve.\label{fig1}}
\end{figure*}
It can be easily verified that the second parenthesis would not be zero for matter fields with $p>-\rho/3$. As a result, one can obtain $A_0$ from the first parenthesis as
\begin{align}\label{AA}
A_0=2\alpha p\left(2H+\f{\dot{H}}{H}\right).
\end{align}
It is interesting to note that if the Universe is filled only by non-relativistic matter with $p=0$, the Lagrange multiplier vanishes on top of FRW Universe.

In the following, we will assume that the Universe is filled by radiation with equation of state $p_r=\rho_r/3$ and dust with equation of state $p_m=0$. As a result, one has $\rho=\rho_m+\rho_r$ and $p=\rho_r/3$, where $r/m$ stand for radiation/dust components, respectively.

Now, define the following set of dimensionless quantities
\begin{align}\label{dimless}
\tau&=H_0t,\quad H=H_0h,\quad \bar{A}_0=H_0A_0,\nonumber\\ \bar{\rho}_i&=\f{\rho_i}{6\kappa^2H_0^2},\quad\Omega_{\Lambda0}=\f{\Lambda}{3H_0^2},\quad\beta=\alpha\kappa^2H_0^4,
\end{align}
where $i=r,m$. From \eqref{cons2}, one can write conservation equations for dust and radiation separately. These equations can then be solved to obtain
\begin{align}\label{cons3}
\bar{\rho}_r=\f{\Omega_{r0}}{a^4},\qquad \bar{\rho}_m=\f{\Omega_{m0}}{a^3},
\end{align}
where $\Omega_{r0}=0.53\times10^{-4}$ and $\Omega_{m0}=0.305$ are present time density parameters for radiation and dust, respectively \cite{plank}. 

Using dimensionless parameters \eqref{dimless} and also equations \eqref{AA} and \eqref{cons3}, one can obtain the Hubble parameter from Friedmann and Raychaudhuri equations as
\begin{align}
h(z)^2=\f{(1+z)\Omega_{m0}+(1+z)^2\Omega_{r0}+\f{\Omega_{\Lambda0}}{(1+z)^2}}{1+4\beta(1+z)^6f(z)},
\end{align}
where we have defined
\begin{align}
f(z)=18\Omega_{m0}^2+45(1+z)\Omega_{m0}\Omega_{r0}+31(1+z)^2\Omega_{r0}^2,
\end{align}
and we have used the redshift variable defined as
\begin{align}
1+z=\f{1}{a}.
\end{align}
In figure \eqref{fig1} we have plotted the Hubble and deceleration parameters, defined as
$$q=(1+z)\f{d \ln h(z)}{dz},$$
for four different values of the model parameter $\beta\times10^{5}=(0.1,0.3,0.6,1.1)$.
It should be noted that in section \ref{pert} we will obtain the best estimation of the parameter $\beta$ using observational data. The best fit value of the parameter is $\beta=0.6\times10^{-5}$ and up to $2\sigma$ confidence level $0.1<\beta<1.11$. As a result the abovee values are chosen to be in the $2\sigma$ confidence interval. We have also plotted the $\Lambda$CDM curve as a red solid curve in these figures. This is obtained by taking  $\beta=0$ in the field equations. The observational data of the Hubble parameter together with their errors is also shown \cite{obshubble}. It can be seen from the figures that the non-minimal coupling between matter and geometry affects the cosmological behavior of the Universe at redshifts greater than $z\sim2$. For the values of the parameter $\beta$ in $2\sigma$ confidence interval, the Universe will decelerate less than standard $\Lambda$CDM theory. Larger values for $\beta$ will make the Universe to accelerate at these redshifts, representing also itself in a smaller values of the Hubble parameter. For negative values of $\beta$, the Universe will have more deceleration at early times, implying that the Universe was larger than the $\Lambda$CDM prediction at that times. We have not plotted this case in figure \eqref{fig1} since these values are not in the $2\sigma$ confidence interval. In summary one could see that the late time observational data would be fulfilled with this non-minimal matter/geometry coupling model. So, more observational evidence would be needed to decide which model will fit the Universe more.
\begin{figure}[H]
	\vspace{0.6cm}
	\includegraphics[scale=0.65]{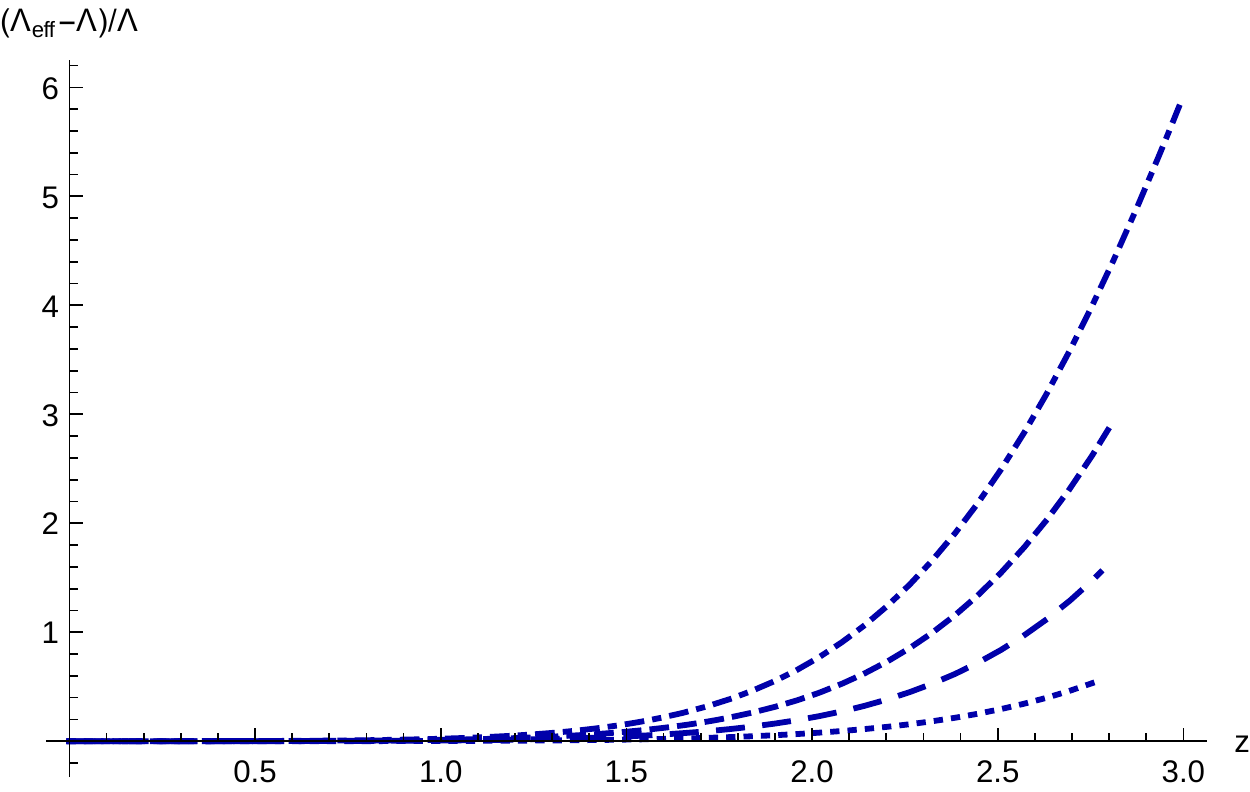}
	\caption{The behavior of the difference between effective cosmological constant with its $\Lambda$CDM value as a function of the redshift $z$ for $\beta\times10^{5}=(0.1,0.3,0.6,1.1)$ indicated as dotted, long-dashed, dashed and dot-dashed respectively.\label{fig2}}
\end{figure}
In figure \eqref{fig2} we have plotted the relative difference between the effective cosmological constant \eqref{coscon} with its $\Lambda$CDM value $\Omega_\Lambda=0.694$ \cite{plank}.  One can see from the figure that the effective cosmological constant becomes equal to the $\Lambda$CDM value at redshifts smaller than $z\sim1$, indicating that the theory becomes identical to the $\Lambda$CDM model. However, the value of the effective cosmological constant differs from the $\Lambda$CDM value for redshifts larger than unity. For values of the parameter $\beta$ in the $2\sigma$ confidence interval we have $\Lambda_{eff}>\Lambda$, implying that the Universe has more acceleration at redshifts $z>1$  with respect to the $\Lambda$CDM model as was discussed before. Also, negative values of the parameter $\beta$ makes the effective cosmological constant to become smaller than its $\Lambda$CDM value and the Universe experiences more deceleration at early times. 
\begin{figure}[H]
	\includegraphics[scale=0.65]{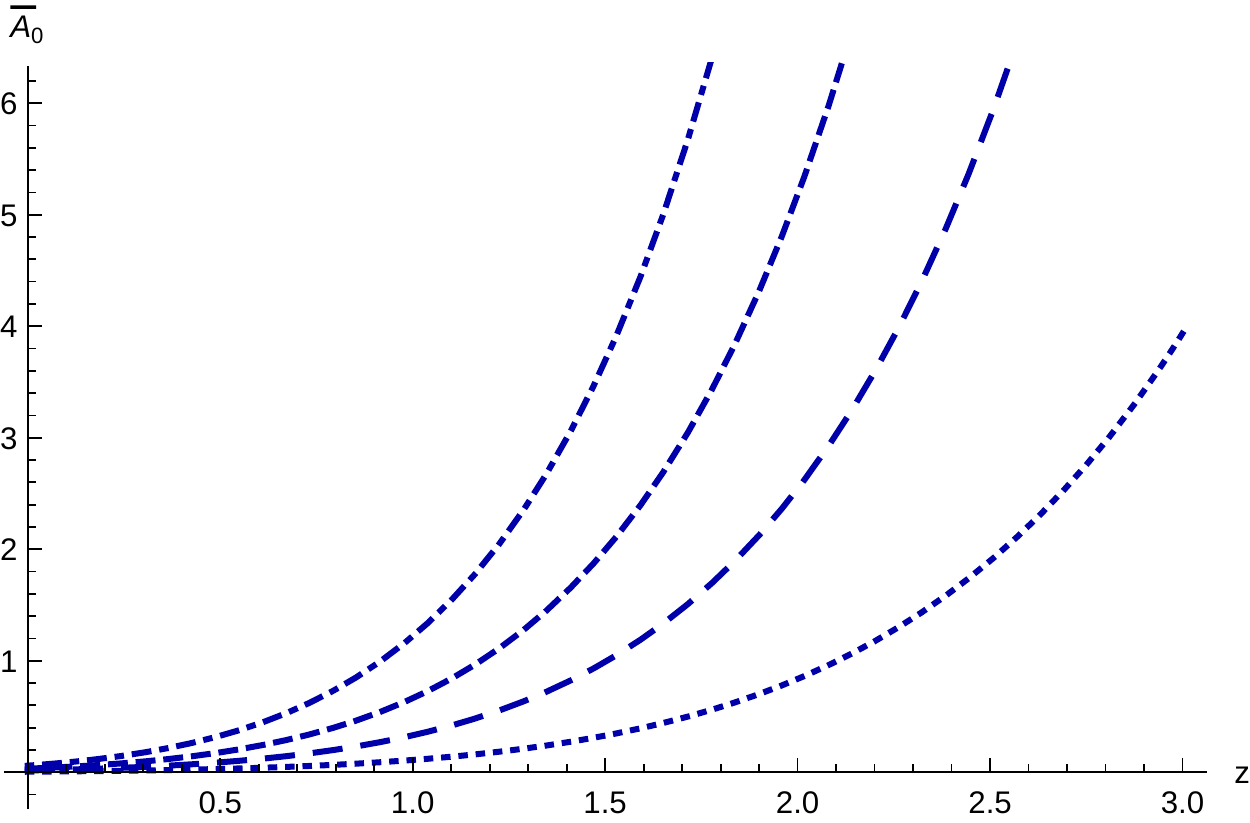}
	\caption{The behavior of the Lagrange multiplier $\bar{A}_0$ as a function of the redshift $z$ for $\beta\times10^{5}=(0.1,0.3,0.6,1.1)$ indicated as dotted, long-dashed, dashed and dot-dashed respectively.\label{fig3}}
\end{figure}
In figure \eqref{fig3}, we have plotted the evolution of the Lagrange multiplier $\bar{A}_0$ as a function of the redshift $z$ using equation \eqref{AA}. One can see from the figure that for the values of parameter $\beta$ in the $2\sigma$ confidence range, the Lagrange multiplier is positive. It should also be noted that for small $\beta$ values, the Lagrange multiplier tends to zero for $z\rightarrow0$. However, for larger values of $\beta$, the Lagrange multiplier remains non-vanishing. This comes back to the fact that as we take the smaller values of the $\beta$ parameter, the theory tends to the standard $\Lambda$CDM model which is conservative. As a result there is no need for an extra force to keep the theory conservative. By letting $\beta$ to adopt larger values, the theory differs significantly from the $\Lambda$CDM theory and one should non-trivially keep it conservative through a Lagrange multiplier. At redshifts larger than unity, as we have discussed earlier, the theory deviates from the $\Lambda$CDM theory and one needs a non-zero Lagrange multiplier for every values of $\beta$.
\begin{figure*}
	\includegraphics[scale=0.38]{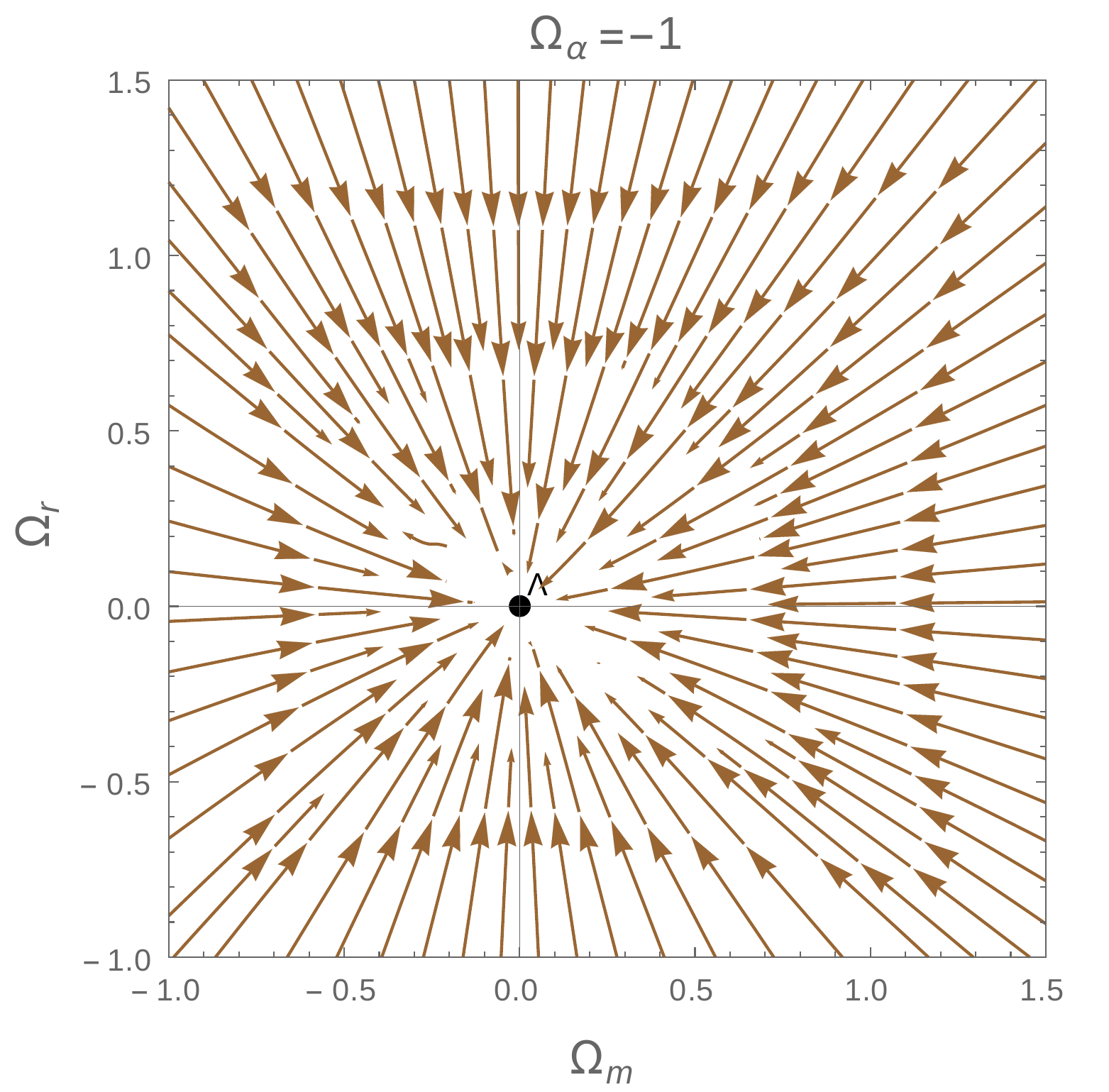}\includegraphics[scale=0.38]{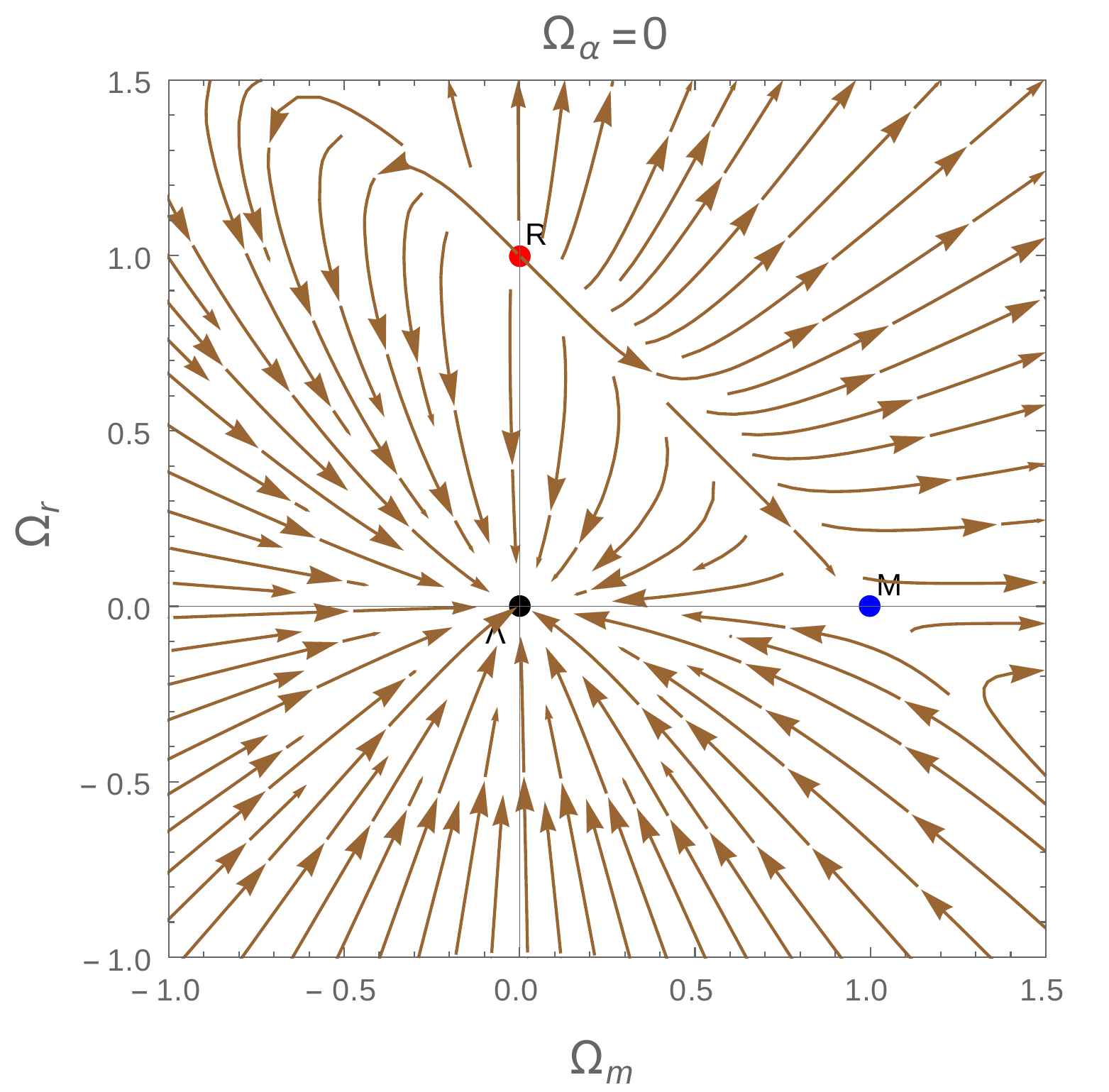}\includegraphics[scale=0.38]{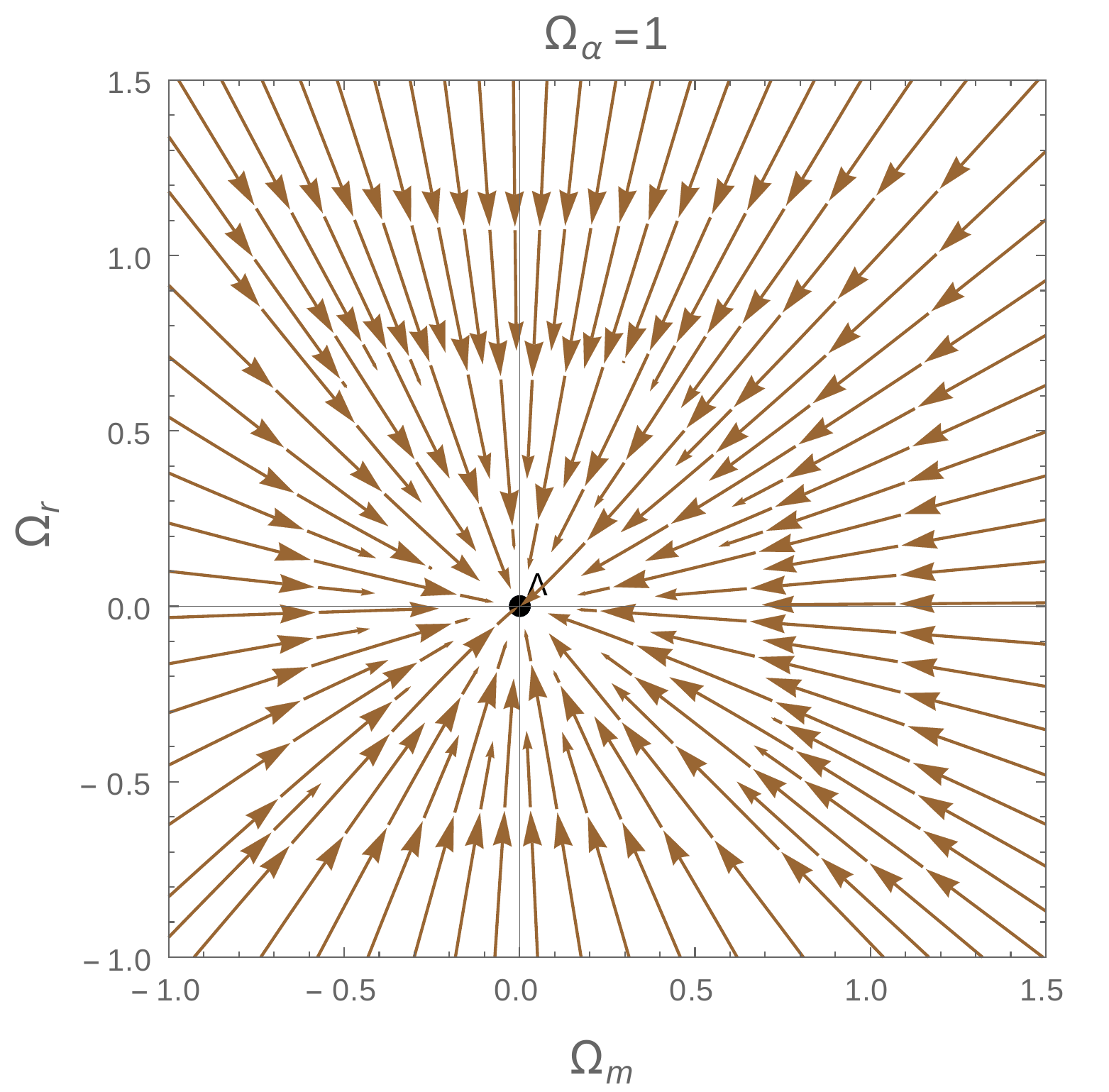}
	\caption{The phase space portrait of the dynamical system \eqref{sys}. We have plotted the $(\Omega_m,\Omega_r)$ planes for three different values of $\Omega_\alpha=-1$ (left), $\Omega_\alpha=0$ (center) and $\Omega_\alpha=1$ (right). The corresponding fixed points are also shown. \label{figsys}}
\end{figure*}
\section{Dynamical system analysis}\label{dyn}
In this section, we will consider the dynamical analysis of the system of equations \eqref{frid1} and \eqref{ray1}. Let us introduce the following set of dynamical variables
\begin{align}
\Omega_m&=\f{\rho_m a^2}{6\kappa^2H^2},\qquad\Omega_r=\f{\rho_r a^2}{6\kappa^2H^2},\nonumber\\
\Omega_\Lambda&=\f{\Lambda a^2}{3H^2},\qquad\quad \Omega_\alpha=\f{\alpha\kappa^2H^4}{a^4}.
\end{align}
It should be noted that the the first three variables are the standard matter density parameters for dust, radiation and the cosmological constant respectively. The last one is related to the non-minimal coupling between matter and geometry which is related to the parameter $\alpha$. Substituting the above variables in the Friedmann equation \eqref{frid1}, one can obtain $\Omega_\Lambda$ as
\begin{align}
\Omega_\Lambda=1-\Omega_m-\Omega_r+\Omega_\alpha(72\,\Omega_m^2+127\,\Omega_r^2+180\,\Omega_m\Omega_r),
\end{align}
where we have used the conservation equations for dust and radiation \eqref{cons3}. It is then evident that the system has three dynamical variables. Using the Raychaudhuri equation \eqref{ray1}, one can obtain the following set of autonomous 3D dynamical system
\begin{subequations}\label{sys}
\begin{align}
\f{d\Omega_m}{dN}&=-A\Omega_m\Big(3+648\,\Omega_\alpha\Omega_m^2+4\,\Omega_r(341\,\Omega_\alpha\Omega_r-1)\nonumber\\&+3\,\Omega_m(660\,\Omega_\alpha\Omega_r-1)\Big),
\end{align}
\begin{align}
\f{d\Omega_r}{dN}&=-A\Omega_r\Big(4+720\,\Omega_\alpha\Omega_m^2+4\,\Omega_r(372\,\Omega_\alpha\Omega_r-1)\nonumber\\&+3\,\Omega_m(660\,\Omega_\alpha\Omega_r-1)\Big),
\end{align}
\begin{align}
\f{d\Omega_\alpha}{dN}&=2A\Omega_\alpha\Big(432\,\Omega_\alpha\Omega_m^2+4\,\Omega_r(248\,\Omega_\alpha\Omega_r-1)\qquad\nonumber\\&+3\,\Omega_m(420\,\Omega_\alpha\Omega_r-1)\Big),
\end{align}
\end{subequations}
where $N=\ln a$ and we have defined
\begin{align}
A=1+4\Omega_\alpha(18\,\Omega_m^2+45\,\Omega_m\Omega_r+31\,\Omega_r^2).
\end{align}
The effective equation of state parameter $\omega_{eff}$ can also be calculated as
\begin{align}
\omega_{eff}&\equiv-\f13-\f23\f{\dot{H}}{H^2}\nonumber\\&
=-\f13A\Big(3+648\,\Omega_\alpha\Omega_m^2+4\,\Omega_r(341\,\Omega_\alpha\Omega_r-1)\nonumber\\&\quad~+3\,\Omega_m(600\,\Omega_\alpha\Omega_r-1)\Big),
\end{align}
The above dynamical system, has three different fixed points which we will discuss in the following.
\subsection{Matter dominated fixed point}
The first fixed point of the system \eqref{sys} is
$$(\Omega_m,\Omega_r,\Omega_\alpha)=(1,0,0),$$
with the effective equation of state parameter $\omega_{eff}=0$. As a result, in this point we have a dust dominated Universe. The eigenvalues of this fixed point are $(-6,3,-1)$, indicating that the dust fixed point is a saddle point.
\subsection{Radiation dominated fixed point}
We also have a fixed point of the system \eqref{sys} corresponding to the radiation dominated phase
$$(\Omega_m,\Omega_r,\Omega_\alpha)=(0,1,0),$$
since the equation of state parameter is $\omega_{eff}=1/3$. The eigenvalues of this fixed point are $(-8,4,1)$, indicating that the radiation fixed point is a saddle point.
\subsection{de Sitter fixed line}
The last fixed point of the system \eqref{sys} is
$$(\Omega_m,\Omega_r,\Omega_\alpha)=(0,0,\Omega_\alpha),$$
which is true for every value of the variable $\Omega_\alpha$. As a result, we have a fixed line with the equation of state parameter $\omega_{eff}=-1$. This corresponds to the de Sitter expansion of the Universe. The eigenvalues of this fixed line is $(-4,-3,0)$, indicating that the de Sitter fixed line is stable. 

In figure \eqref{figsys}, we have plotted the $(\Omega_m,\Omega_r)$ phase space planes for three different values $\Omega_\alpha=-1,0,1$. We have also shown the fixed points in the figures. It should be noted that the dust and radiation fixed points are located at $\Omega_\alpha=0$ plane, which can be seen in the second plot. The de Sitter line is the line perpendicular to the planes and crosses the $(0,0)$ point. As one can see from the figures, the de Sitter fixed line is an attractor and all lines will end at this fixed point eventually. It is interesting to note that if one starts from the radiation dominated fixed point there is a flow which let the Universe to transform to dust dominated phase and then continue its evolution to the de Sitter stable fixed point. The theory can then in principle has a history such that the radiation and matter dominated phases occurs before the Universe falls into the late time de Sitter evolution. As a result the thermal history of the Universe can be recovered in this model. 
\section{Matter density perturbations}\label{pert}
Let us consider the first order scalar perturbations of the model \eqref{eq1}. The perturbed metric can be written in the Newtonian gauge as
\begin{align}
ds^2=a^2(t)\Big[-(1+2\varphi)dt^2+(1-2\psi)d\vec{x}^2\Big],
\end{align}
where $\varphi$ and $\psi$ are the Bardeen potentials.
The perturbed energy momentum tensor can also be written as
\begin{align}
&\delta T^0_0=-\delta\rho\equiv-\rho\delta,\nonumber\\& \delta T^0_i=(1+w)\rho\partial_i v,\quad\nonumber\\& \delta T^i_j=\delta^i_jc_s^2\rho\delta.
\end{align}
In the above expression, we have defined the matter density contrast $\delta$ as $\delta=\delta\rho/\rho$. Also $v$ is the scalar mode of the velocity perturbation associated with the matter sector, $c_s$ is the sound speed and $w$ is the equation of state parameter of baryonic matter $p=w\rho$.
In the following, we will assume that the Universe at the time where we are performing the perturbations is in the dust dominated phase. As a result, the perturbed and unperturbed matter content of the Universe satisfy $\delta p/\delta\rho=c_s^2=p/\rho$, where $c_s^2=0$.

For the vector field $A_\mu$ we define the first order scalar perturbation as
\begin{align}
\delta A_\mu=a(\mc{A}_0,-\partial_i\mc{A}).
\end{align}
Perturbing the conservation equation \eqref{cons} to first order in perturbation variables gives
\begin{align}\label{pert1}
\dot\delta+\left(3H+\f{\dot\rho}{\rho}\right)(\delta-2\varphi)-3\dot\psi+\theta=0,
\end{align}
and
\begin{align}\label{pert2}
\dot\theta+H\theta-k^2\varphi=0,
\end{align}
where we have defined $\theta=\nabla_i\nabla^i v$, and we transform to the Fourier coordinates, where $\vec{k}$ is the wave vector. Using the above equations and also the background field equation \eqref{cons2}, one obtains a differential equation of the evolution of the matter density contrast as
\begin{align}\label{pertmain}
\ddot\delta+H\dot\delta+k^2\varphi-3H\dot\psi-3\ddot\psi=0.
\end{align}
Let us consider the sub-horizon limit of the theory where $k\gg H$. In this limit the $(00)$ component of the metric field equation and the $(0)$ component of vector field equation at first order in perturbations reduces to
\begin{align}\label{pert3}
2\alpha H\rho^2\dot\delta&+(a^2\rho+4\alpha k^2\rho^2-24\alpha H^2\rho^2)\delta\nonumber\\&+k^2a\rho\mc{A}+4\kappa^2 k^2\psi=0,
\end{align}
\begin{align}\label{pert4}
a\mc{A}+\alpha\rho(\varphi+3\psi)=0.
\end{align}
Also, the $i\neq j$ components of the metric field equation \eqref{eq1} reads
\begin{align}\label{per5}
\kappa^2(\varphi-\psi)=a\rho\mc{A}.
\end{align}
From equations \eqref{pert3}-\eqref{per5}. one can obtain the perturbation variables $\varphi$, $\psi$ and $\mc{A}$ as
\begin{align}\label{pert6}
\psi=\f{\kappa^2+\alpha\rho^2}{\kappa^2-3\alpha\rho^2}\varphi,\quad\mc{A}=-\f{4\alpha\kappa^2\rho}{a(\kappa^2-3\alpha\rho^2)}\varphi.
\end{align}
and
\begin{align}\label{pert7}
k^2\varphi=&-\f{\rho(\kappa^2-3\alpha\rho^2)}{4\kappa^4}\Big[2\alpha H\rho\dot\delta\nonumber\\&+(a^2+4\alpha k2\rho-24\alpha H^2\rho)\delta\Big].
\end{align}
Using equation \eqref{pert7} in \eqref{pertmain}, one obtains for the evolution of the matter density contrast
\begin{align}
\ddot\delta&+\Big[1-\f{\alpha}{2\kappa^2}(\kappa^2-3\alpha\rho^2)\rho^2\Big]H\dot\delta\nonumber\\&
-\f{\kappa^2-3\alpha\rho^2}{4\kappa^4}\Big[a^2+4\alpha(\kappa^2-6H^2)\rho\Big]\rho\delta=0.
\end{align}
It should be noted that in the case $\alpha=0$, the above equation reduces to the standard equation in $\Lambda$CDM theory. One can see from the above equation that the presence of the non-minimal matter/geometry coupling affects the effective gravitational constant and also the friction term in the evolution equation of the density contrast.

Transforming to dimensionless variables \eqref{dimless}, noting that the quantity $\delta$ is dimensionless by its own and using the expression \eqref{cons3} for the dust energy density $\rho_m$, one obtains
\begin{widetext}
\begin{align}\label{eqdelta}
\delta^{\prime\prime}&+\left[\f{h^\prime}{h}+18\beta\,\Omega_{m0}^2(1+z)^5\left(1-108\beta\,\Omega_{m0}^2(1+z)^6\right)\right]\delta^\prime\nonumber\\&-\f32\f{\Omega_{m0}}{(1+z)h^2}\Big(1-108\beta\,\Omega_{m0}^2(1+z)^6\Big)\Big(1+24\beta\,\Omega_{m0}(1+z)^5(\gamma^2-6h^2)\Big)\delta=0,
\end{align}
\end{widetext}
where we have transformed to the redshift coordinates 
and prime denotes derivative with respect to the redshift $z$. It should be noted that the evolution equation of the matter density contrast depends on $\gamma=k/H_0$.

In order to solve the above equation, we will use the same initial conditions as in $\Lambda$CDM theory in deep matter dominated era in which
\begin{align}
\f{d\delta}{d\ln a}|_{z_\star}=\delta|_{z_\star},
\end{align}
where $z_\star$ is some point in the deep matter dominated era which we will assume to be $z_\star=7.1$.

In order to compare the model with observational data, we will use the data set on the Hubble parameter in the redshift range $0<z<2$ \cite{obshubble} and also the observational data on $f\sigma_8$ \cite{fsigma8}
\begin{align}
f \sigma_8 \equiv \sigma_8(z)f(z),
\end{align}
where $\sigma_8(z)=\sigma_8^0\, \delta(z)/\delta(0)$, and $\sigma_8^0$ is a model dependent constant. We have defined the growth rate of matter perturbations as
\begin{align}
f=\f{d \ln\delta}{d\ln a}=-(1+z)\f{\delta'}{\delta}.
\end{align}
We estimate the values of $\sigma_8^0$ and also the model parameter $\beta$ by maximizing the Likelihood function
\begin{align}
\mathcal{L}=\mathcal{L}_{0}e^{-\chi^2/2},
\end{align}
where $\mathcal{L}_0$ is the normalization constant and the quantity $\chi^2$ in our case  is given by
\begin{align}
\chi^2&=\chi_H^2+\chi_{f\sigma_8}^2\nonumber\\&=\sum_i\left(\f{H_{i,obser}-H_{i,theory}}{\sigma_i}\right)^2\nonumber\\&~~~~+\sum_j\left(\f{f\sigma_{8j,obser}-f\sigma_{8j,theory}}{\sigma_j}\right)^2,
\end{align}
where  $H_{i,theory}$ and  $f\sigma_{8j,theory}$ are the theoretical values  for the observables $H_{i,obser}$ and $f\sigma_{8j,obser}$ and $\sigma_i$ is the error of the $i$th data. It should be noted that the two sets of data on $H$ and $f\sigma_8$ that we used in this paper are independent, so the total Likelihood function is obtained by multiplying the individual Likelihoods of the two sets.

\begin{table}
	\begin{center}
		\begin{tabular}{|c||c||c||c|} 
			\hline
			$~~~~~$& ~~best fit~~&~~~~~~$1\sigma$ interval~~~~~~&~~~~~~$2\sigma$ interval~~~~~~ \\
			\hline
			\hline
			$\beta\times10^5$ &0.6&$0.3<\beta<0.8$ & $0.1<\beta<1.11$   \\
			\hline
			$\sigma_8^0$ &0.71& $0.69<\sigma_8^0<0.73$ & $0.68<\sigma_8^0<0.74$ \\
			\hline
		\end{tabular}
	\end{center}
	\caption{The best fit values together with their $1\sigma$ and $2\sigma$ confidence intervals of the model parameter $\beta$ and also for $\sigma_8^0$.\label{tab1}}
\end{table}

In table \eqref{tab1} we have summarized the best fit values of the parameters $\beta$ and $\sigma_8^0$ and their $1\sigma$ and $2\sigma$ confidence intervals. It should be noted that for $\sigma_8^0$ at the best fit point, we have $\chi^2/dof=0.44$.

\begin{figure}[H]
	\includegraphics[scale=0.5]{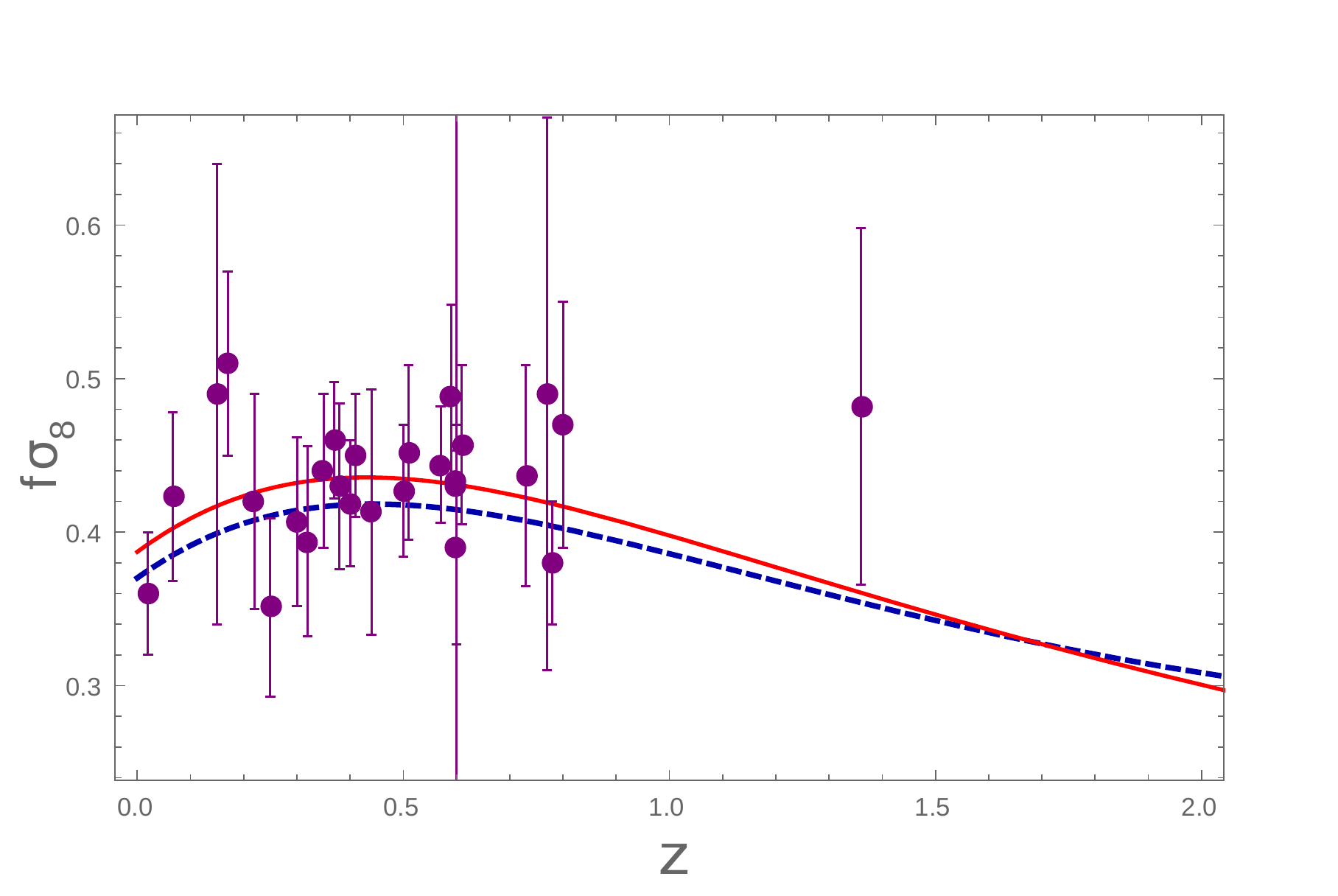}
	\caption{The evolution of the $f\sigma_8$ as a function of redshift $z$. The Solid curves correspond to the $\Lambda$CDM theory and the dashed line is associated with our model with $\beta=0.6\times10^{-5}$ and $\gamma=1.4$. We have also plotted the observational data with their errors in the figure. \label{figfs8}}
\end{figure}

In figure \eqref{figfs8}, we have plotted the evolution of the $f\sigma_8$ as a function of redshift $z$ for the best fit values shown in table \eqref{tab1} and $\gamma=1.4$.

One can see from figure \eqref{figfs8} that the qualitative behavior of the $f\sigma_8$ function is similar to the $\Lambda$CDM model. For small redshift values, one can see that the value of the $f\sigma_8$ is less than that of $\Lambda$CDM theory. This implies that this model predicts slower growth rate of matter fields. However, as an observational side, both theories are satisfactory and so, more observations would be needed to favor one of them.
\section{Conclusions and final remarks}\label{conc}
In this paper, we have considered a modified theory of gravity containing non-minimal coupling between matter and geometry. The new term can be considered as a generalization of the energy-momentum squared gravity by coupling it to the Ricci tensor. As, we have a non-minimal coupling between matter fields and gravity, the energy-momentum tensor is no longer conserved. We have fixed this problem by adding a constraint through Lagrange multiplier to ensure that the equation $\nabla_\mu T^{\mu\nu}=0$ holds. The Lagrange multiplier in fact is a vector field and can be considered as a potential to keep the matter components to behave conservative. In the FRW background we have plotted the Lagrange multiplier in figure \eqref{fig3}. In this figure, it can be seen that the Lagrange multiplier is always positive, which means that we should do positive work to keep the theory conservative. Also, the Lagrange multiplier tends to zero at the present time. Closer we are to the early times, stronger force we should imply to make the theory conservative. Moreover, the covariant divergence of the metric field equation becomes the dynamical equation for the Lagrange multiplier. In the FRW Universe, this equation becomes an algebraic equation for the Lagrange multiplier and one can obtain the Lagrange multiplier as a function of pressure and the Hubble parameter \eqref{AA}. This confirms that the source for the conservative force is in fact the non-minimal nature of the matter/geometry coupling itself. In the case of dust dominated Universe, the Lagrange multiplier vanishes. This is true only at the background level since at the first order perturbation level, the Lagrange multiplier is always non-zero unless the energy density vanishes.

In this paper, we have found the best fit values for the model paramter $\alpha$ and also for $\sigma_8^0$ by using two sets of independent data corresponding to the Hubble parameter and also the $f\sigma_8$ function. In order to satisfy observational data, the best fit value of the model paramter $\alpha$ and also its 2$\sigma$ confidence interval shows that this parameter should be small and positive. In this range, the behavior of the Hubble parameter coincides with the $\Lambda$CDM theory for redshifts smaller than $z\sim2$. However for larger redshifts the Hubble parameter becomes smaller. This shows that the present theory predicts larger Hubble radius at that times compared to the $\Lambda$CDM model. The analysis of the deceleration parameter shows that the Universe at early times could be in the accelerating phase for large enough values of the parameter $\alpha$. For the best fit value of $\alpha$, however, the behavior of the Universe at early times is qualitatively the same as $\Lambda$CDM theory, despite the fact that we have less deceleration in the present theory.
\begin{figure}[H]
	\vspace{0.6cm}
	\includegraphics[scale=0.53]{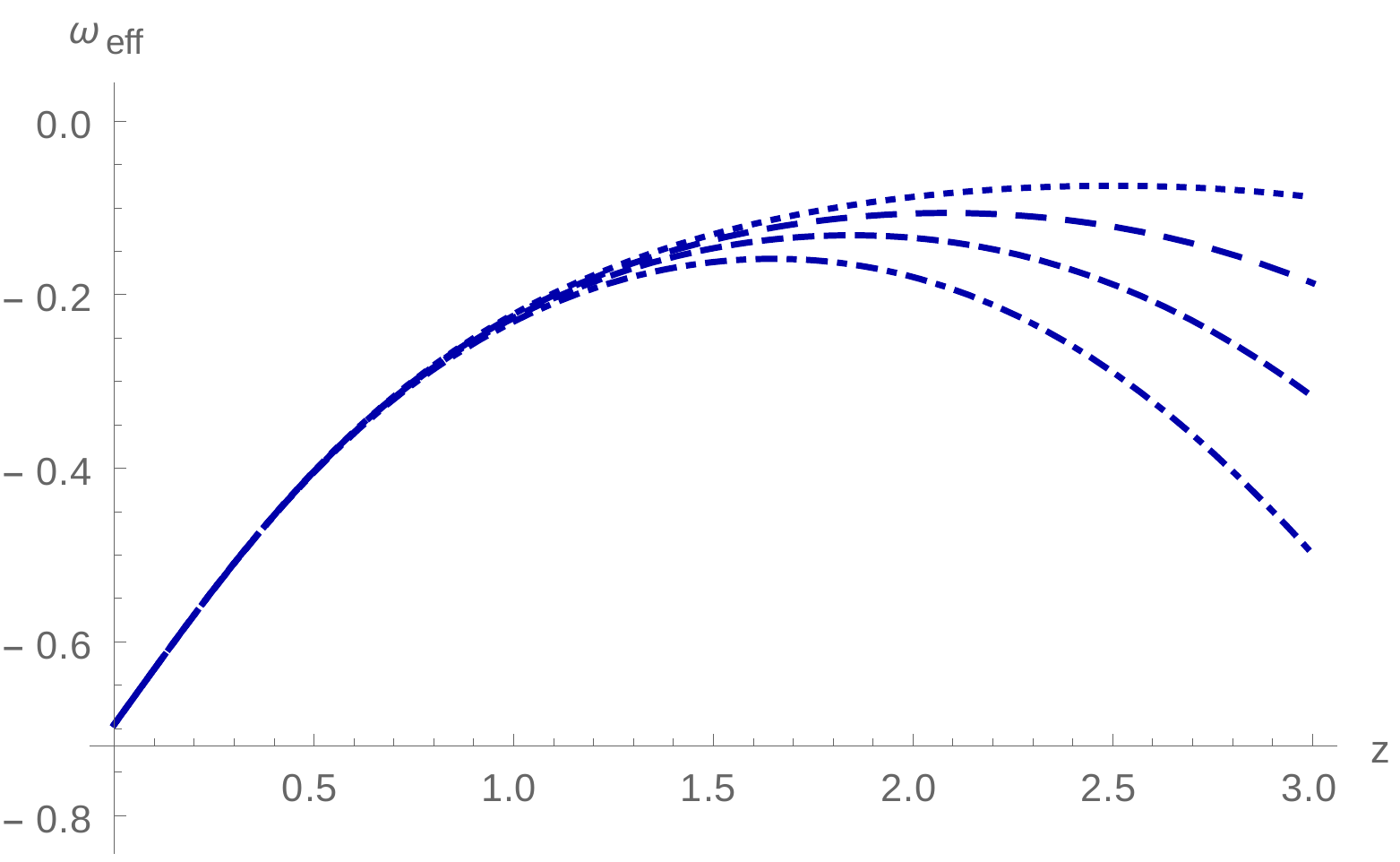}
	\caption{The behavior of the effective equation of state parameter as a function of the redshift $z$ for $\beta\times10^{5}=(0.1,0.3,0.6,1.1)$ indicated as dotted, long-dashed, dashed and dot-dashed respectively.\label{figomega}}
\end{figure}
It is well-known that in the semi-classical regime, the vacuum expectation value of the energy-momentum tensor renormalizes the cosmological constant $\Lambda$, as well as the gravitational constant $G$. In this paper, we have considered a non-minimal coupling between matter and geometry through the interaction of the Ricci tensor with the energy-momentum tensor squared. In the semi-classical regime this new coupling would contribute to the renormalization of the coefficients of the Ricci tensor and also to derivatives of the Riemann tensor. Remembering the definition of the effective cosmological constant \eqref{coscon}, one expects that quantum corrections to the matter fields, renormalizes $\Lambda_{eff}$ as well. It should also be noted that the effective energy-momentum tensor does not alter the exact moment of phase transitions at early times. This is because the theory is constrained to have a conservation of the energy-momentum tensor as we have implied by introducing a Lagrange multiplier to the action. However, the behavior of the matter abundances in radiation, dust and dark energy dominated phases differ from the standard $\Lambda$CDM model.

We have also analyzed the dynamical evolution of the model. The theory is a three dimensional autonomous system. We have shown that the theory has three fixed points similar to $\Lambda$CDM theory, corresponding to the dust, radiation and de Sitter expansion. However, in the present model, the fixed point corresponding to the de Sitter expansion is changed to a fixed line which is the $\Omega_\alpha$ axes in the phase diagram of the theory. However, in the $\Omega_\alpha=0$ plane the qualitative behavior of the theory is the same as in the $\Lambda$CDM theory.

We have also considered the first order perturbation analysis of the theory. Since we have a non-minimal matter-geometry coupling, the anisotrpic stress $\eta=(\varphi-\psi)/\varphi$ is non-vanishing. From equation \eqref{eqdelta}, one can see that the non-minimal coupling adds some new terms proportional to at least $(1+z)^5$ to the evolution equation of the density contrast. This shows that the matter/geometry coupling would be important for large values of $z$. For small redshifts, these terms are negligible and the qualitative behavior of the density contrast is similar to the $\Lambda$CDM theory. This can also be seen from the plot of $f\sigma_8$ in figure \eqref{figfs8} where one can see that the late time observational data could be satisfied in this theory. It should be noted that the value of $f\sigma_8$ is smaller than the corresponding value of $\Lambda$CDM mode.

Let us note about the $H_0$ and $\sigma_8$ tensions in this model. In general, dynamical dark energy theories with $\omega<-1$ could in principle loosen the $H_0$ tension but they worsen the $\sigma_8$ tension. Conversely, theories with $\omega>-1$ does not change the status of $H_0$ tension but they loosen the $\sigma_8$ tension. In figure \eqref{figomega}, we have plotted the $\omega_{eff}$ as a function of the red-shift $z$ for $\beta\times10^{5}=(0.1,0.3,0.6,1.1)$. One can see from the figure that our model belongs to the sub-class of dynamical dark-energy models with $\omega>-1$. As a result we expect that the theory would make the $H_0$ tension better but consequently will make the $\sigma_8$ tension worse. However, more detailed analysis would be needed to investigate this point.

At last, we would like to note that the non-minimal coupling between matter and geometry could in principle explains late time cosmology, but more analysis is needed to choose the best theory among many late time cosmological models.

\end{document}